\documentclass[12pt,letterpaper]{article}

\usepackage{graphicx,subfigure,fullpage,natbib}
\bibpunct{(}{)}{;}{a}{}{,}
\begin{document}

\begin{titlepage}
\begin{center}

\section*{The Importance of Hands-on Experience with Telescopes for Students}

\bigskip

George C. Privon\footnotemark[1]\\
(434) 924-7491 --- \texttt{gcp8y@virginia.edu}\\
\medskip
Rachael L. Beaton\footnotemark[1], David G. Whelan\footnotemark[1], Abel Yang\footnotemark[1], Kelsey Johnson\footnotemark[1], Jim Condon\footnotemark[2]\\
\medskip

\footnotetext[1]{University of Virginia}
\footnotetext[2]{National Radio Astronomy Observatory -- Charlottesville}

\end{center}

\section*{Abstract}

Proper interpretation and understanding of astronomical data requires good knowledge of the data acquisition process. The increase in remote observing, queue observing, and the availability of large archived data products risk insulating astronomers from the telescope, potentially reducing their familiarity with the observational techniques crucial in understanding the data. Learning fundamental observing techniques can be done in at least three ways:
\begin{itemize}
  \item{College and university operated observing facilities}
  \item{Student involvement in national facilities through competitive proposals}
  \item{Programs at national facilities to increase upper-level undergraduate and graduate student exposure to telescopes} 
\end{itemize}
We encourage both national organizations and universities to include programs and funding aimed at supporting hands-on experience with telescopes through the three methods mentioned.
\end{titlepage} 

\section{Introduction}

Recent technological developments have fundamentally altered the traditional model of astronomical observation.  As the CCD replaced the photographic plate,  the astronomer moved out of the dome and into an often separate control room. As instruments have become more complex and their basic operation requiring more skill, the presence of instrument operators have reduced the role of the astronomer in the on-site data acquisition. With the advent of queue and remote observing, the astronomer need not leave their office, let alone home institution, to acquire their desired data. This shift from the traditional model of astronomical observations certainly has its advantages in terms of the more effective use of monetary resources, reduction of travel time and making data more accessible to the astronomical community. Additionally, an increasing amount of science is being conducted using archival data or data from large systematic surveys (e.g. the Sloan Digital Sky Survey).

However, there are downsides to the decrease in direct interaction with astronomical instruments and facilities. The growing use of archives to mine for data has the effect of separating students from the problems associated with observing such as the importance of instrumental uncertainties in data sets, knowledge of instrument construction, and the propagation of errors in reduction pipelines. An understanding of the inherent problems associated with observing is critical to pass on to students, for whom the experience gained by using telescopes and becoming familiar with observational techniques can enhance their understanding of data sets (from their own observations or from archives), and the scientific implications of errors in their data.

\section{Importance of Student Exposure to Telescopes}

Large data archives provided by survey telescopes (such as the Sloan Digital Sky Survey) and long-term telescope projects (like the Hubble and Spitzer Space Telescopes) provide great opportunities for quickly building large data sets of objects for scientific analysis \citep[e.g. satellite debris streams, see][]{2006ApJ...642L.137B}. Students limited to mining such archives risk having a narrow scope of knowledge about the telescopes instruments being used, since the focus of the project is primarily on the analysis. In this position, the students risk losing insight into fundamental observational techniques, including: target selection, photometry and/or spectroscopy methods, astrometry, and error analysis. An understanding of all of the steps of the astronomical research process is necessary to correctly interpret \citep[e.g.,][]{Sivakoff05}. Even more crucial is an understanding of how each of these aspects complement the other. Systematic effects can be important factors when interpreting data from surveys \citep[e.g.,][]{Coleman90,Rahmani09}. Students who learn how to treat systematic effects through first hand experience have the ability to apply their skills to any telescope system, and can properly account for systematic errors in their analysis.

In contrast with large data archives, queue observing can still leave astronomers needing to work with raw (or pipeline processed) data. While less removed from the data acquisition than using survey data, queue observing still insulates observers from the telescope. However, the need to understand the data acquisition steps will still remain vital to proper interpretation. Queue observing can result in more efficient completion of observing projects \citep{Boroson96}, and will undoubtedly become more prevalent in the future. However, the need to understand the data acquisition steps will remain as important as ever.

\section{Ensuring Student Exposure to Research Facilities}

Gaining knowledge of observational techniques can be fairly simple. A night or two of observing at even a small telescope can make concepts more concrete than a semester of lectures. Therefore observing, even at small facilities, is the best way to learn observational techniques. The hands-on experience gained at telescopes, including an introduction to instrumentation and the development of analysis software are unquestionably important tools in astronomy. 

We wish to emphasize the importance of student observing runs at telescope facilities of all types. Hands-on experience with observing will provide a foundation that is necessary when designing the next generation of instruments and data analysis software. For example, the increasing use of large surveys and archival data sets is making pipeline processing more ubiquitous. This makes the necessity for an accurate reduction pipeline even stronger. Only with a suitable understanding of the limitations in the data acquisition process can one expect to construct an accurate data pipeline, properly tweaked to maximize data potential. In cases where the raw data is not or cannot be archived, constructing an accurate pipeline is even more crucial because re-running the pipeline on old data is not an option \citep[see][for a look at the impossibility of archiving the data to be taken with the Large Synoptic Survey Telescope]{Becla06}.

It is therefore important that we ensure adequate opportunities for students to learn observational techniques in order to better design an instrument, observatory or survey that is cost-effective and scientifically useful. This will require an understanding of the scientific principles behind a question and the various techniques and procedures that may be used to obtain scientifically useful data. Hands-on experience at an observing facility will enhance their understanding of the details of both data acquisition and analysis, and will train the next generation of astronomers to be as conscious of the limitations of their observations as the present generation.

\section {Methods of Exposure}

Ensuring student exposure to telescopes can be achieved through at least three main avenues, with overlap between them:
\begin{enumerate}
  \item{Colleges and universities utilizing their own facilities}
  \item{Telescope access on national facilities through the competitive proposal process}
  \item{Programs at national facilities to support student exposure}
\end{enumerate}

Some colleges and universities operate their own observatories or are partners in observatory collaborations. These two types of facilities likely allow the most convenient means of exposing students to an observatory and to the data collection process. It also permits a greater degree of experimentation and risk taking. Additionally, faculty mentoring greatly enhances the insight students gain. Who better to explain how an instrument works than those who use it most? 

Student exposure to telescopes does not need to be in the context of a world-class telescope. In fact, due to the high expenses and goal for efficiency, these facilities can be the wrong place for training. Students exhibit improved understanding of astronomical concepts through hands-on learning with even modest exposure to small scale facilities \citep[e.g.][]{Jacobi08}. Many important principles can be taught through the use of a small optical or radio telescope. Even a $12$" optical telescope with a CCD camera can be used to explore basic principles of optical observing. A small radio telescope, such as those developed by the MIT Haystack Observatory\footnote{http://www.haystack.mit.edu/edu/undergrad/srt/}, can introduce basic radio astronomy. This class of facilities is conducive to student exploration and experimentation.

The second method to increase student exposure is through competitive proposals to national facilities. Owing to the competitive nature, this would likely involve more student-faculty interaction and would involve single students rather than groups of students working as a team. The advantage here is it provides opportunity to work on larger facilities that are typical in professional astronomy. However, the importance of hands-on learning remains central as queue observing only exposes students to a portion of the full experience.

The third avenue is by programs through national facilities and missions offering opportunities to teach students about instrumentation and observing. REU programs have been successful in teaching students how to do science with existing data. While sometimes involving data reduction, they do not always involve students in the data acquisition aspect of a project. We encourage the NSF to expand the REU program to include first- and second-year graduate students, giving them the same opportunities to gain familiarity with observational techniques.

Funding allocated to the EPO budgets is generally focused towards K-12 education and the general public. Extending this focus to the upper level undergraduate and graduate student demographic can enhance their knowledge of the data acquisition, reduction, and analysis process, as suggested here. Including undergraduate and graduate students in national EPO planning (i.e. NSF or AAS) would provide support for university departments, teaching courses in observational techniques.

Funding from EPO budgets at national facilities could also be used to increase the number of ``summer school'' style conferences. If aimed at upper level undergraduate and graduate students they could provide a beneficial avenue for exposing them to observational techniques. The advantage of ground based national facilities is both their coverage of different wavelengths (e.g. radio, optical) and their accessibility to students.

A common aspect of all three options mentioned is the simple potential to inspire. This is just as important as the technical knowledge. The inspiration can come in the form of the thrill of an instrumental challenge, the insight from a scientific observation, or even the simple impact of seeing a large telescope for the first time. Indeed, \citet{Slater08} notes: ``It is generally accepted that most undergraduate non–science majors taking an astronomy course spend little to no time actually looking through a telescope.'' Early experience with a telescope can encourage students to take up a career in astronomy or a related field. We have included an appendix containing personal experiences from the graduate students on this paper which demonstrate the inspiration that can occur.

\section{Conclusion}

Working knowledge of astronomical telescopes is important to both observers and theorists. Knowledge of the limitations inherent in each data set are critical to a proper interpretation. The most effective way of learning about these limitations is through hands-on exposure in observational situations.

To ensure a suitable level of competence in this area, large university and national observing projects should include programs and funding in their EPO budgets aimed specifically at graduate and upper-level undergraduate students. Additionally, we encourage colleges and universities to support their own smaller observatory programs to advance student learning. It can act as a springboard to empower students to effectively use world-class instruments creatively and efficiently.

\clearpage

\appendix

\section{Graduate Student Experiences}

We include below, personal experiences from the four graduate students on this paper to provide anecdotal examples of the value and benefit of ensuring direct participation in the data acquisition process.

\subsection{Rachael L. Beaton}

As an undergraduate, I aggressively pursued astronomy and began working with my undergraduate adviser as a first year student. It was not until my third year as an undergraduate that I obtained my first observing experience on a world class telescope, and in fact on a telescope of any size. At that point in my career, I was enrolled in an observational methods class, actively memorizing or deriving equations related to telescopes and instruments. I had already reduced two large imaging projects and was actively working with the 2MASS 6x survey of M31. I already had a blossoming research career, but had never used a telescope myself. Though I was certain that astronomy was my chosen profession, it was not until I had that first observing experience that I really felt the type of overwhelming motivation to devote myself to the field.

I distinctly remember seeing the 4-meter at Kitt Peak National Observatory for the first time. I remember identifying the various components of the telescope and for the first time truly understanding how they work together. It was only the first night of observing that my adviser realized that I was attacking my research and the observing campaign with a new level of fervor. He chuckled as I looked over every pixel of every new image, in awe at how excited I was with each distant galaxy I stumbled upon. He was laughing at me as he joined me in my review of the images. Shortly thereafter, laughing ceased, he also joined me in my child-like excitement. Soon after, I abandoned my homework and he abandoned the paper he was working on and together we began on-the-fly reducing the night's data to assess its quality. We did extensive literature searches, compared our instrumental CMDs to published data and began making proactive adjustments to our program. In the morning, utterly exhausted, my adviser remarked that it had been a long time since he had felt so moved by data acquisition and that in viewing the world through the eyes of his student, seeing the telescope for the first time, he had renewed his own astronomical inspirations. Instead of feeling weighted by the remaining four nights of our run, he said he felt as invigorated as I.

Four years later and as a graduate student, I have transitioned to working more heavily with queue observatories (LBT) and with remote observing (APO). Every time I conceive a project for these facilities, I rely heavily on my observational background, the hands on experience that I have gained through thirty-one nights of observing on site at 4-meter and larger class telescopes. Having worked with other students who have not had the same opportunities with large telescopes, I can attest that my experience gives me the insight to more quickly diagnose observational problems while observing and to better understand every level of data reduction, most specifically in understanding both the limitations and potential of my data. As a Teaching Assistant in both introductory and upper level undergraduate observing courses, I have now been inspired by the excitement of my students as they acquire their first data on our teaching facilities at UVa (ranging from 8-inch Meades with eyepieces to spectroscopy and photometry with a 40-inch telescope). I have witnessed the changes in their conceptualization of astronomy and in their further maturation as young scientists. I find that this hands-on experience has been as crucial for them as it was for me. As a result of my training, I view all research projects as a form of art and as with all art, its mastery is found through a combination of continued inspiration and through hard fought experience. I sincerely hope that the hands on aspects of training new astronomers is forever preserved. 

\subsection{George C. Privon}

My first two experiences with telescopes in a ``serious'' astronomical fashion occurred while an undergraduate at the Rochester Institute of Technology. After expressing my interest in astronomy to a faulty member (Michael Richmond), he got me involved in the monitoring of cataclysmic variable stars. I used a 12" telescope and a cooled CCD camera to do time series photometry in concert with various amateur astronomers around with world, all part of the Center for Backyard Astrophysics. It was a simple, yet effective introduction to observing and was my first exposure to the methodology I have continued to use in my research of Active Galactic Nuclei today.

Also while at RIT I (along with one other undergraduate student) had the great opportunity to help a faculty member (Andrew Robinson) with an observing run using the Mayall 4m at Kitt Peak National Observatory. I can still remember walking out of the control room and seeing the telescope for the first time. It was very inspiring to stand so close to such a large telescope. Even more so knowing that later that night we would be opening the dome to use it. Unfortunately we took no data due to poor weather, but the trip was a great experience that certainly inspired me to continue pursuing a career in astronomy.

\subsection{David G. Whelan}

I am a second-year graduate student at the University of Virginia. I first took an interest in astronomy while attending Ithaca College, majoring in physics.It was the fall of my senior year that I observed at the Palomar 200" telescope, my first observing run at a major telescope facility. Seeing such a large telescope in operation was 'the nail in the coffin,' sealing my interest and motivation in the subject. Since then, I have worked extensively with data from ground- and space-based observatories, and spent two very fruitful years at Cornell before coming to UVa, where I helped write reduction software for the Spitzer/IRS. A new-found interest in modeling is fortified by my observing experience: the models I use of super star clusters have to be compared to actual observations, so placing my star cluster models at astronomical distance with consideration of telescope field of view, spatial and spectral resolution is of the utmost importance.

\subsection{Abel Yang}

As a theoretician, I am not actively involved in data acquisition or most aspects of observation. Nevertheless, the experience of having visited a number of research facilities, and having used a research grade telescope has had a positive impact on my research.

Since theory which is not backed up by observations is little more than a mathematical exercise, on many occasions I have been on the lookout for possible surveys that will test the theories I am working on. Visiting the SDSS telescope and using its neighbor the ARC 3.5m at Apache Point, I have not only gained an understanding of the process through which data is obtained but more importantly the strengths and limitations that arise from each step of the process. Together with a physical understanding of the universe, this knowledge has come in useful when assessing and working with survey data products.

For example, in the course of my research, the GOODS spectroscopic redshift catalog was used by another team to obtain the counts in cells distribution of galaxies at high redshifts. Knowing the limitations of the instrument(in this case a multi-object slit spectrograph) meant that collisions may pose a problem and that the catalog might not be suitable for such a purpose. An alternative technique(used by the COSMOS survey) would be to use high resolution multi-band imaging with well-calibrated photometric redshifts instead. This would have solved the collision problem provided each object was identified properly, but would have greater uncertainties in the measured redshifts, possibly leading to a considerable amount of Eddington bias.

This example demonstrates that a knowledge of observational techniques complements physical intuition in selecting a suitable data set to address a specific question. I feel that in general, this sets the direction of my research to one which produces testable predictions, and enables the effective use of survey data products to test these predictions. In addition, I believe this background may also be useful in designing future surveys.

\end{document}